\documentclass[11pt]{article}
\usepackage{amsmath,amssymb,color,cite}
\usepackage{graphicx}
\usepackage{authblk}
\usepackage{amsfonts}
\usepackage{enumerate}
\usepackage{bbm}
\usepackage{multirow}
\usepackage{nicefrac}
\usepackage[all]{xy}
\usepackage{graphicx}
\usepackage{booktabs}
\usepackage{hyperref}
\usepackage{amsfonts}

\def\n{\hat{n}}

\def\s{{\sigma}}
\def\b{{\beta}}
\def\r{{\rho}}
\def\l{{\lambda}}

\newcommand{\hoch}[1]{$\, ^{#1}$}

\makeatletter
\@addtoreset{equation}{section}
\makeatother

\newcommand{\be}{\begin{equation}}
\newcommand{\ee}{\end{equation}}
\newcommand{\bea}{\setlength\arraycolsep{2pt} \begin{eqnarray}}
\newcommand{\eea}{\end{eqnarray}}
\newcommand{\nn}{\nonumber}

\def\bea{\begin{eqnarray}}
\def\eea{\end{eqnarray}}
\newcommand{\ft}[2]{{\textstyle\frac{#1}{#2}}}
\def\a{\alpha}
\def\b{\beta}
\def\g{\gamma}
\def\G{\Gamma}
\def\d{\delta}

\def\e{\epsilon}

\def\k{\kappa}
\def\l{\lambda}

\def\m{\mu}
\def\n{\nu}
\def\r{\rho}
\def\s{\sigma}

\def\t{\tau}

\def\o{\omega}

\begin{document}
\begin{flushright}
\hfill{ \
\ \ \ \ }
\end{flushright}

\vspace{25pt}
\begin{center}
{\Large {\bf Attractor Mechanism and Non-Renormalization Theorem in 6D (1,0) Supergravity}
}

\vspace{30pt}

{\Large
Yi Pang
}

\vspace{10pt}

\hoch{1}{\it Center for Joint Quantum Studies and Department of Physics,\\
School of Science, Tianjin University, Tianjin 300350, China\\
and Mathematical Institute, University of Oxford, \\
Woodstock Road, Oxford OX2 6GG, U.K. \vskip 5pt }

\vspace{10pt}

\vspace{20pt}

\underline{ABSTRACT}
\end{center}
\vspace{15pt}
We compute the macroscopic entropy of the supersymmetric rotating dyonic strings carrying linear momentum in 6D (1,0) supergravity with curvature squared corrections. Our calculation is based on Sen's entropy function formalism applied to the near-horizon geometry of the string solution taking the form of an extremal BTZ$\times S^3$. The final entropy formula states that the two independent supersymmetric completions of Riemann tensor squared contribute equally to the entropy. A further $S^3$ compactification of the 6D theory results in a matter coupled 3D supergravity model in which the quantization condition of the SU(2)$_R$ Chern-Simons level implies the horizon value of the dilaton is not modified by higher derivative interactions  beyond supersymmetric curvature squared terms.

\thispagestyle{empty}

\pagebreak
\voffset=-40pt
\setcounter{page}{1}

\tableofcontents


\section{Introduction}
The most striking feature hidden in the black hole thermodynamics 
is the area law of the entropy instead of the usual  volume law observed in nearly all local physical systems. Efforts devoted to understanding such an area law of entropy have led to the discovery of the holographic principle of quantum gravity realized concretely in the fruitful framework of AdS/CFT correspondence. One precursor of AdS$_3/$CFT$_2$ correspondence is precisely the microstates counting of the BPS black holes \cite{SV1} in string theory. In the infinite charge limit, the logarithm of  the corresponding 2D CFT density of states successfully recovers the Bekenstein-Hawking formula. Subsequent studies have extended the agreement between the macroscopic and microscopic entropy to large classes of BPS black holes \cite{BMPV,MSW}  in string/M theory (see \cite{Peet1, David:2002wn, sen1,Mandalsen} for nice reviews).

When the charges are large but finite, the matching between the macroscopic and microscopic entropy implies deviations from the area law that are suppressed by inverse powers of the large charges\footnote{There can also be logarithmic corrections to the area law. However, 
they are not present in ${\cal N}=4$ theory \cite{Sen3}, which is the relevant context of this paper.}. Astonishingly, for a class of 4D static  BPS black holes, it has been demonstrated \cite{CdM1,CdM2} that the supergravity action supplemented by higher derivative terms does provide the right amount of deviations needed to  match with its CFT counterpart. Specifically, up to leading-order higher derivative corrections, the macroscopic entropy of the 4D black holes agrees precisely with the CFT result computed in \cite{MSW,Vafa:1997gr}
\begin{equation*}
\label{cfte}
S_{\rm micro}=2\pi\sqrt{\ft16\hat{q}_0\left(C_{ABC}\hat{p}^A\hat{p}^B\hat{p}^C+c_{2A}\hat{p}^A\right)}\,,
\end{equation*}
where the coefficients $C_{ABC}$ and $c_{2A}$ encode the geometric data about the internal manifold on which string/M-theory is compactified. In particular, $c_{2A}$ labels the coefficients in front of the leading higher derivative terms in the low-energy effective action. Comparison between the BPS black hole entropy and Eq. \eqref{cfte} relies on an implicit map from the conserved charges carried by the black hole to the CFT data $\hat{q}_0$ and $\hat{p}^A$. At leading order, this identification is straightforward. However, when higher derivative interactions are switched on, the relation between the black hole charges and $(\hat{q}_0$, $\hat{p}^A)$ becomes less obvious and may be modified by higher derivative interactions as pointed out in \cite{CDKL,dK} for a class of 5D spinning BPS black holes. Supersymmetric higher derivative terms in $D>4$ introduce another subtlety--the non-gauge invariant terms like $A_{(1)}\wedge{\rm Tr}( R\wedge R)$ in $D=5$ and $B_{(2)}\wedge {\rm Tr}(R\wedge R)$ in $D=6$ which require a careful treatment regarding their contributions to the macroscopic entropy. In fact, \cite{CDKL,dK} have both computed the macroscopic entropy for the same class of spinning BPS black holes with curvature squared corrections. However, their results match only in the static limit.

For a subclass of black holes considered in \cite{CDKL,dK}, computing higher derivative corrections to the macroscopic entropy can be performed in a different setup. Specifically, this subclass of BPS black holes corresponds to M-theory branes wrapped around cycles in K3$\times T^2$, which amount to strings or branes in IIA string theory wrapped on K3$\times S^1$. In the latter case, the leading $\a'$ corrections arise from string one-loop effects \cite{LM}. In supersymmetric  compactifications of string/M-theory, one can first retain only the lower-dimensional supergravity multiplet first and subsequently incorporate  matter multiplets. Specific to IIA string theory compactified on K3, the 10D string frame action reduces to an effective action of the form $e^{-2\phi}R+R^2+\cdots$ where the $R^2$ term descends from the $R^4$ term in $D=10$. Supersymmetrizing this type of action can be readily performed utilizing the dilaton Weyl multiplet within the framework of off-shell ${\cal N}=2$ supergravity. The resulting supersymmetric $R^2$ action is relatively simple, independent of the dilaton field.  On the other hand, compactifications of 11D supergravity on Calabi-Yau threefolds leads to a 5D supergravity model of the schematic form $R+f(\phi) R^2+\cdots$. In this case, the 5D standard Weyl multiplet is the preferable building block. Rescaling the metric in either case inevitably complicates the curvature squared superinvariants. Therefore, neatness of the higher derivative action selects the suitable off-shell supergravity multiplets. There is the view that in higher derivative theories, one can always perform field redefinitions under which the structure of the higher derivative terms are not preserved while the physics remains the same. Therefore, the specific form of the higher derivative terms is not crucial. However, this also means that one is free to choose a frame in which the calculation can be done most efficiently. As we will see later, such a convenient frame for studying supersymmetric black holes naturally comes from the off-shell formulation of supergravity, without which the field equations will be much more complicated. 

In this work, we shall revisit the BPS black hole entropy in the IIA setup for reasons below. It is recalled that the K3 compactification of IIA string preserves 16 supercharges. The supersymmetric curvature squared action consistent with the same amount of supersymmetry was obtained only recently in 6D off-shell supergravity (and its dimensional reduction) based on the dilaton Weyl multiplet \cite{NOPT,BNOPT}, where the invariance under eight supercharges is manifest. A further $S^1$ reduction of the 6D model gives rise to the aforementioned 5D model
arising from IIA string on K3$\times S^1$. In practice, we shall uplift the 5D black hole solutions along the $S^1$ direction to six dimensions and readdress the issue of the macroscopic entropy there. We emphasize that it is totally equivalent to calculate the entropy of black holes in the 5D model based on the dilaton Weyl multiplet as circle reduction preserves conserved quantities carried by the solution. We prefer the 6D model for computational simplicity as the 6D supergravity multiplet consists of fewer fields.  In the 6D setup, we will also explore whether different supersymmetric completions of the Riemann tensor squared give the same contribution to the black hole entropy, by virtue of the fact that there is a unique off-shell formulation of 6D (1,0) supergravity allowing for well-defined two- and higher derivative superinvariants \cite{Bergshoeff:1985mz,NOPT,BNOPT} (all curvature squared superinvariants based on the 5D dilaton Weyl multiplet were previously obtained in \cite{BRS,OP1,OP2})\footnote{There exists also a standard Weyl multiplet in $D=6$. However, the two-derivative supergravity action based on the standard Weyl multiplet is pathological for the lack of a saddle point \cite{Bergshoeff:1985mz}.}. 
This is advantageous to the 5D setup based on standard Weyl multiplet \cite{CDKL,dK}, where only one supersymmetric completion of Riemann tensor squared preserving eight supercharges \cite{HOT} instead of 16 supercharges is known explicitly. Up to date, there exists no rigorous proof that the other independent supersymmetric completion of the Riemann tensor squared based on standard Weyl multiplet, if one exists, contributes equally to the BPS black hole entropy. 

In the next section, we introduce the 6D (1,0) supergravity action equipped with
curvature squared superinvariants. For simplicity, we restrict our discussions to NS-NS sector fields that are captured by the dilaton Weyl multiplet. Upon circle reduction, the 6D model reduces to the 5D ungauged STU model extended by curvature squared terms. Without higher derivative corrections, black hole solutions in the 5D STU model are well studied \cite{CY,CC}. Among them, we single out the three charge rotating black holes obtained in \cite{CY} and lift them to 6D rotating dyonic strings with linear momentum. The BPS limit of the 6D strings requires equal angular momenta. As a consequence, the solution exhibits an enhanced U(2)$\times\mathbb{R}^2$ isometry, based on which the general ansatz underlying the BPS string solution and compatible with the 6D off-shell supersymmetry can be derived.  Following a procedure proposed in \cite{CP}, one can then apply the ansatz to construct rotating  BPS dyonic string solutions with higher derivative corrections. The entropy of the string solution is determined from its near-horizon geometry in the form of an extremal BTZ$\times S^3$. In Sec. III 
by extremizing the entropy function defined for the near-horizon geometry of BPS string solutions in $R^2$ extended 6D (1,0) supergravity, we are able to express the entropy of the string solution in terms of the near horizion charges $q_i$ which are conjugate variables of the parameters appearing in the near-horizon profile of the string. However, in order to compare with existing results in the literature, it is necessary to identify the near-horizon charges $q_i$ in terms of conserved charges measured at infinity. In our case, these are the charges characterizing the number of fundamental strings or wrapped NS5 branes, angular momentum of the string and the linear momentum excitation. 
The main task of Sec. IV is to find the relation between near-horizon charges and global charges by solving the asymptotically flat half BPS string solutions and zooming into its near-horizon region. Combining the results from Sec. III and IV, we obtain the following interesting results i) the size of the extremal BTZ horizon does receive corrections although previous work  \cite{NOPT,BNOPT} has verified that the $AdS_3$ and $S^3$
radii are not modified by the curvature squared superinvariants; ii) our entropy formula recovers precisely the result obtained in \cite{dK} specialized to the K3 compactification; iii) the two independent supersymmetric completions of the Riemann tensor squared contribute equally to the BPS string entropy. We conclude in Sec. V with discussions emphasizing the nonrenormalization of the horizon value of the dilaton beyond the leading $\a'$ correction. 

\section{The model}
In this section, we set up the 6D supergravity model describing the low-energy dynamics of 
IIA string on K3 with leading $\a'$ corrections that are generated by string one-loop effects. In the low-energy effective action, the massless bosonic fields consist of a metric, an antisymmetric two-form, 24 vectors and 81 scalars. In the framework of 6D (1,0) off-shell supersymmetry, the metric, antisymmetric two-form and one scalar denoted as $\{g_{\m\n}\,,B_{\m\n}\,,L\}$ belong to the bosonic sector of the dilaton Weyl multiplet. Henceforth, we will call $L$ the dilaton field which is related to the string coupling via $g_s^2=1/\langle L\rangle_{\rm vev} $. The black hole solutions to be considered later are fully captured by this field content. We thus write out terms in the effective action composed by the dilaton Weyl multiplet. Vectors and hypermultiplet scalars can be consistently truncated out from the field equations. The leading term in the action takes the form   
\be
\label{LEH}
{\cal L}_{\rm EH}=\sqrt{-g}L\left(R+L^{-2}\partial_{\mu}L\partial^{\mu}L- \ft{1}{12}H_{\m\n\r}H^{\m\n\r} \right)\,.
\ee
As explained in \cite{BNOPT}, the string frame action given above can be coupled to curvature squared superinvariants without modifying the supersymmetry transformation rule. There are three independent curvature squared superinvariants corresponding to the supersymmetrization of the pure Riemann tensor squared \cite{BSS86}
\be
\label{Riemsq}
{\cal L}_{{\rm Riem}^2}= \sqrt{-g}\left[ R_{\m\n\a\b}(\o_-) R^{\m\n\a\b}(\o_-)- \ft14 \e^{\m\n\r\s\l\t} B_{\m\n} R_{\r\s}{}^{\a}{}_{\b} (\o_-) R_{\l\t}{}^{\b}{}_{\a}(\o_-)\right],
\ee
the supersymmetric completion of the Gauss-Bonnet combination 
\bea
\label{GBSUGRA}
{\cal L}_{\rm GB}&=&\sqrt{-g} \left[  R_{\m\n\r\s} R^{\m\n\r\s} - 4 R_{\m\n} R^{\m\n} + R^2 + \ft16 R H^2 - R^{\m\n} H_{\m\n}^2 + \ft12 R_{\m\n\r\s} H^{\m\n\l} H^{\r\s}{}_{\l} \right. \nn\\
&&\qquad \left. + \ft{5}{24} H^4 + \ft{1}{144} (H^2)^2 - \ft18 (H_{\m\n}^2)^2 - \ft14 \e^{\m\n\r\s\l\t} B_{\m\n} R_{\r\s}{}^{\a}{}_{\b} (\o_+) R_{\l\t}{}^{\b}{}_{\a}(\o_+) \right],
\eea
and the supersymmetric Ricci scalar squared term whose explicit form can be found in \cite{O, BNOPT}. One can verify that field equations derived from the supersymmetric Ricci scalar squared are proportional to the those from the leading order action \eqref{LEH} and, thus, does not modify the solution at the order we are interested in. Moreover, its contribution to the BPS string entropy vanishes \cite{CP}. Therefore the Ricci scalar squared superinvariant will be omitted from later discussions. In the above supersymmetric curvature squared actions, $R_{\m\n\r\s}$ is the standard Riemann tensor of the metric, while $R_{\m\n}{}^{\a}{}_{\b} (\o_\pm)$ is the curvature defined with respect to the torsionful spin connections $\o_{\pm\m}^{\a}{}_{\b}$,
\be
R_{\m\n}{}^{\a}{}_{\b} (\o_\pm)=\partial_{\m}\o_{\pm\n}^{\a}{}_{\b}+\o_{\pm\m}^{\a}{}_{\g}\,\o_{\pm\n}^{\g}{}_{\b}-(\m\leftrightarrow\n)\,,\quad \o_{\pm\m}^\a{}_\b=\o_{\m}^\a{}_\b\pm\ft12H_{\m}{}^\a{}_\b\,.
\ee
The shorthand notations for various contractions of $H_{\m\n\r}$ are defined as
\be
H^2=H_{\m\n\r}H^{\m\n\r}\,,\quad H_{\m\n}^2=H_{\m\r\s}H_{\n}{}^{\r\s}\,,\quad H^4=H_{\m\n\s}H_{\r\l}{}^{\s}H^{\m\r\d}H^{\n\l}{}_{\d}\,.
\ee

In summary, we will study the entropy of BPS black holes using the model
\be
\label{IIAK3}
S_{R+R^2}=\frac1{16\pi G_6}\int d^6x\sqrt{-g}\left({\cal L}_{\rm EH}+\frac{\l_1}{16}\,{\cal L}_{{\rm Riem}^2}+\frac{\l_2}{16}\,{\cal L}_{\rm GB}\right)\,,
\ee
where the combination with $\l_1=\l_2=\a'$ descends from IIA compactified on K3 and enjoys  a supersymmetry enhancement. For general values of $\l$, the effective action \eqref{IIAK3} preserves only eight supercharges. In later discussions, we keep $\l$ general in order to tell apart the contribution to the BPS string entropy from the individual curvature squared superinvariant. 

\section{Macroscopic entropy from extremizing entropy function}

The 6D rotating dyonic string \cite{CL} that comes from lifting the 5D 3-charge rotating BPS black hole \cite{CY} can be put in the form
with a manifest U(2)$\times\mathbb{R}^2$ isometry
\bea
\label{genansatz}
ds_6^2&=&-a_1^2(r)(dt+\varpi\s_3)^2+a_2^2(r)(dz+A_{(1)})^2+b(r)^2dr^2+\ft14c^2(r)(\s_3^2+d\theta^2+\sin^2\theta d\phi^2)\,,\nn\\
B_{(2)}&=&2P\o_2+d(r)dt\wedge dz+f_1(r)dt\wedge\s_3+f_2(r)dz\wedge\s_3\,,\quad A_{(1)}=A_0(r)dt+A_3(r)\s_3\,,\nn\\
L&=& L(r)\,,
\eea
where in our notation $\s_3=d\psi-\cos\theta d\phi$ and $d\o_2={\rm Vol}(S^3)$. The $r$-dependent functions in \eqref{genansatz} are given by
\bea
a_1^2&=&\frac{r^4}{(r^2+Q_1)(r^2+Q_2)}\,,~a_2^2=\frac{r^2+Q_2}{r^2+Q_1}\,,~ c^2=r^2+P\,,~ b^2=c^2 r^{-2}\,,~ \varpi=\frac{\n}{2r^2}\,,\nn\\
d&=&\frac{r^2}{r^2+Q_1}\,,~ A_0=\frac{r^2}{r^2+Q_2}\,,~f_1=0\,,~ f_2=-d\varpi\,,~ A_3=A_0\varpi\,,~ L=\frac{r^2}{d(r^2+P)}\,.
\eea
In the expression above,  
$Q_1$ characterizes the number of fundamental string while $Q_2$ denotes the linear momentum excitations of the string along the $z$ direction.

The near-horizon limit is attained by zooming in the $r=0$ region while keeping other parameters fixed. In terms of the new coordinates and parameter defined below, 
\be
\r=r^2\,,\quad \tau= \frac{2t}{\sqrt{P}Q_1\r_+}\,,\quad \widetilde{\psi}=\psi+\frac{2\n}{PQ_1}z\,,\quad \r^2_+=\frac{Q_2}{Q_1}\left(1-\frac{\n^2}{PQ_1Q_2}\right)
\ee
the near-horizon geometry can be expressed as an extremal BTZ$\times S^3$
\bea
ds^2_{6,{H}}&=&\frac{P}4(-\r^2d\tau^2+\frac{d\r^2}{\r^2})+{\r}^2_+(dz+\frac{\sqrt{P}}{2\r_+}\r d\tau)^2+\frac{P}4(\widetilde{\s}_3^2+d\theta^2+\sin^2\theta d\phi^2)\,,\nn\\
B_{(2),{H}}&=&-\frac{P}4\cos\theta d\phi\wedge d\widetilde{\psi}+\frac{\sqrt{P}\r_+}{2}\r d\tau\wedge dz-\frac{\n}{2Q_1}dz\wedge d\psi\,,\quad L_{ H}=\frac{Q_1}P\,,
\eea
in which $\widetilde{\s}_3=d\widetilde{\psi}-\cos\theta d\phi$ and evidently the last term in $B_{(2),H}$ is a pure gauge. 
Locally, the near-horizon geometry preserves eight supercharges \cite{GMR,AP,KuLu} while the asymptotically flat string solution preserves four supercharges. 

In the following, we will adopt the entropy function formalism \cite{sen1} to compute the macroscopic entropy of the BPS string solution with higher derivative corrections. The attractor mechanism was initially proposed to compute entropy of extremal black holes in 4D ${\cal N}=2$ supergravities. The procedure summarized here can be readily generalized to $AdS_3$ by recasting the $AdS_3$ as U(1) bundled over $AdS_2$ as we shall do later (see Eq. \eqref{3dansatz}). Since the near-horizon geometry of a regular extremal black hole is locally $AdS_2$, it can be parametrized as
\be
\label{geo}
g_{\mu\nu}dx^{\mu}dx^{\nu}=v(-\rho^2d\tau^2+\frac{d\rho^2}{\rho^2})\,,\quad F^{(i)}_{\rho\tau}=e_i\,,\quad \phi_i=u_i\,,
\ee
where ``$i$" labels the different vector multiplets and the explicit dependence on magnetic charges has been suppressed. The attractor mechanism states that the near-horizon values $\vec{u}\,,\vec{q}$ and $v$ can be obtained by extremizing
the so-called entropy function
\be
{\cal E}(\vec{u},\vec{q},v)=2\pi(\vec{e}\cdot\vec{q}-f(\vec{u},\vec{q},v))
\ee
at fixed $q_i$, where $f(\vec{u},\vec{q},v)$ is the action evaluated on the geometry \eqref{geo} and $q_i$ are the electric charges conjugate to the near-horizon variables $e_i$. Notice that we do not introduce conjugate variables for magnetic charges in the entropy function. The reason is that the variational principle associated with the original action is achieved by fixing the electric potential and magnetic charge.

The near-horizon geometry of the BPS string solution is a direct product of BTZ with $S^3$, we first reduce the 6D theory to $D=3$. For the purpose of this work, we adopt the the minimal ansatz 
\be
d\hat{s}_6^2=ds_3^2+\frac{P}4(\s_3^2+d\theta^2+\sin^2\theta d\phi^2)\,,\quad\hat{H}_{(3)}=dB_{(2)}+\frac{P}4\sin\theta d\theta\wedge d\phi \wedge d\psi\,,\quad \hat{L}=L\,,
\ee
which suffices to capture the near-horizon geometry of the BPS string solution even with higher derivative corrections. The fields on the rhs of the equations above are independent of the coordinates on $S^3$. Parameter $P$ labels the $S^3$ radius squared and sets up the physical scale relative to which one can discuss corrections to the leading-order solution, as the 6D model does not have an intrinsic cosmological constant.  The dimension reduction of leading two-derivative Lagrangian yields \footnote{In \cite{Deger:2014ofa}, it was shown that a more complete consistent $S^3$ reduction of the two-derivative 6D supergravity coupled to a single chiral tensor multiplet leads to the SO(4)$\ltimes\mathbb{R}^6$ gauged ${\cal N}=4$ supergravity in $D=3$. Based on this result, some black string solutions in $D=6$ were obtained by uplifting solutions in the 3D theory \cite{Deger:2019jtl}.}
\be
\label{3dEH}
{
\cal L}^{(3)}_{\rm EH}=\sqrt{-g}L(R-L^{-2}\partial_{\mu}L\partial^{\mu}L-\frac{1}{12}|H_{(3)}|^2+\frac4P)\,,\quad H_{(3)}=dB_{(2)}\,.
\ee
It is straightforward to reduce the supersymmetric curvature squared actions  and the results are 
\bea
\label{3daction}
{\cal L}^{(3)}_{{\rm Riem}^2}&=&\sqrt{-g}\left( 4R_{\m\n}R^{\m\n}-R^2-\ft13H_{(3)}\Box H_{(3)}-\ft16R |H_{(3)}|^2+\ft1{48}(|H_{(3)}|^2)^2\right)+L_{{\rm CS}}(\o_-)\,,\nn\\
{\cal L}^{(3)}_{\rm GB}&=&\frac{16}{P}\sqrt{-g}(R+\ft1{12}|H_{(3)}|^2)+L_{\rm CS}(\o_+)\,,\nn\\
L_{\rm CS}(\o_-)&=&\frac{8}{\sqrt{P}}\left(-\ft12{\rm tr}(\G d\G+\ft23 \G^3)-\ft1{24}H_{(3)}|H_{(3)}|^2+\ft12H_{(3)} R\right)\,,\nn\\
L_{\rm CS}(\o_+)&=&\frac{8}{\sqrt{P}}\left(-\ft12{\rm tr}(\G d\G+\ft23 \G^3)+\ft1{24}H_{(3)}|H_{(3)}|^2-\ft12 H_{(3)}R\right)\,,
\eea
where the parity odd terms are given as three-forms. Next we parameterize the putative metric and two-form
\be
\label{3dansatz}
ds_3^2=\frac{\ell^2}4\left(-\r^2dt^2+\frac{d\r^2}{\r^2}\right)+r_+^2\left(dz+\frac{\ell \r}{2r_+}d\t\right)^2\,,\quad r_+=\frac{\ell}{2e_1}\,,\quad B_{(2)}=e_2 \r d\t\wedge dz\,,
\ee
where we choose $\ell>0$ in our convention. Together with the constant dilaton $L:=L_H$, we have four independent variables $\{\ell\,,e_1\,,e_2\,,L_H\}$. Off-shell supersymmetry implies a model-independent relation between $e_1$ and $e_2$
\be
\label{susye}
4e_1e_2=\ell^2\,,
\ee
which should be employed after varying the entropy function with respect to $\{\ell\,,e_1\,,e_2\,,L_H\}$. One should also be aware that there is no similar off-shell relation between $e_2$ and $L_H$. Although one of the BPS equations relates $L$ to $d$, $e_2$ is defined with respect to time $\t$ differing from the original time $t$ by a numerical factor that depends on the detail of the solution.

The entropy function is the Legendre transform of the effective action for the reduced phase space variables $\{\ell\,,e_1\,,e_2\,,L_H\}$. It is constructed as follows. We first parameterize a generic three-dimensional field configuration with a Killing vector $\partial_z$ as 
\be
\label{redansatz}
\hat{g}_{MN}=g_{mn}(y)dy^mdy^n+\phi(y)\left(dz+C_m(y) dy^m\right)^2\,,~ \hat{L}=L(y)\,,~ \hat{B}_{(2)}=b_m(y)dy^m\wedge dz\,,
\ee
where we have used a ``hat" to label three-dimensional fields and $m\,,n=1,2$. Note that a term of the form $b(y)dy^1\wedge dy^2$ in $B_{(2)}$ has been omitted, since it does not contribute to the field strength $H_{(3)}$ which appears in the Lagrangian. 
In terms of the two-dimensional fields, the action associated with \eqref{3dEH} and \eqref{3daction} can be expressed as
\be
S=\int d^2x\left[\sqrt{-{\rm det}g}({\cal L}_0^{(2)}+{\cal L}_1^{(2)})+{\cal L}_2^{(2)}\right]\,,
\ee
where $\sqrt{-{\rm det}g}{\cal L}_{0}^{(2)}$ denotes the circle reduction of the two-derivative Lagrangian density \eqref{3dEH} and $\sqrt{-{\rm det}g}{\cal L}_{1}^{(2)}$ is the circle reduction of the diffeomorphism invariant part of the combined  higher derivative Lagrangian density $\frac{1}{16}(\l_1{\cal L}^{(3)}_{{\rm Riem}^2}+\l_2{\cal L}^{(3)}_{\rm GB})$. They are obtained by substituting \eqref{redansatz} into the corresponding three-dimensional Lagrangian densities. Thus 
\be
\sqrt{-{\rm det}g}\left[{\cal L}_0^{(2)}+{\cal L}_1^{(2)}\right]=\int dz\sqrt{-{\rm det}\hat{g}}\left[{\cal L}_0^{(3)}+{\cal L}_1^{(3)}\right]=2\pi\sqrt{-{\rm det}\hat{g}}\left[{\cal L}_0^{(3)}+{\cal L}_1^{(3)}\right]\,.
\ee
The remaining two-dimensional Lagrangian density ${\cal L}_2^{(2)}$ is the dimensional reduction of 
the pure Lorentz Chern-Simons term in the combined higher derivative Lagrangian above. It is given in \cite{SS} and takes the form
\be
{\cal L}_2^{(2)}
=\frac{K}2\pi\left[R\varepsilon^{mn}F_{mn}+\varepsilon^{mn}F_{mp}F^{pq}F_{qn}\right]\,,\quad K=-\frac{\l_1+\l_2}{32\pi\sqrt{P}G_3}\,,
\ee
where $F_{mn}=\partial_mC_n-\partial_nC_m$ and $\varepsilon^{01}=1$.
Following \cite{sen1}, we define
\be
f(\ell,e_1,e_2,L_H)=\sqrt{-{\rm det}g}\left[{\cal L}_0^{(2)}+{\cal L}_1^{(2)}\right]+{\cal L}_2^{(2)}\,,
\ee
evaluated on the ansatz \eqref{3dansatz}.
Its explicit form is given by
\bea
&&f(\ell,e_1,e_2,L_H)=
\frac{1}{8G_3}\Big[L_H\left(\frac{4e_1e_2^2}{\ell^3}+\frac{\ell^3}{2e_1P}-\frac{3\ell}{4e_1}\right)+\l_1\left(\frac{24e_1^3e_2^4}{\ell^9}-\frac{3e_1e_2^2}{\ell^5}+\frac{3}{32e_1\ell}
\right.\nn\\
&&\left.+\frac{8e_1^2e_2^3}{\ell^6\sqrt{P}}-\frac{3e_2}{2\ell^2\sqrt{P}}-\frac{1}{4e_1\sqrt{P}}\right)+\l_2\left(\frac{3e_2}{2\ell^2\sqrt{P}}-\frac{8e_1^2e_2^3}{\ell^6\sqrt{p}}-\frac{3\ell}{4e_1P}-\frac{4e_1e_2^2}{\ell^3P}-\frac{1}{4e_1\sqrt{P}}\right)\Big]\,.\nn\\
\eea
Then the entropy function is defined as \cite{sen1} 
\be
\label{sfunction}
{\cal E}=2\pi\left(e_1q_1+e_2q_2-f(\ell,e_1,e_2,L_H)\right)\,,
\ee
where $q_1$ and $q_2$ are conserved charges conjugate to $e_1$ and $e_2$ respectively. It should be remarked that although the asymptotically flat solution carries 4 conserved charges, however, in the near-horizon geometry, only two of them are electric . This can be seen by compactifying the near-horizon geometry of the BPS string solution on a 2-torus. 

In order to compute the black hole entropy, we must extremize the entropy function for fixed $q_1$ and $q_2$. This is achieved by solving 
\be
\label{exme}
\frac{\partial{\cal E}}{\partial \ell}=0\,,\quad \frac{\partial{\cal E}}{\partial e_1}=0\,,\quad \frac{\partial{\cal E}}{\partial e_2}=0\,,\quad \frac{\partial{\cal E}}{\partial L_H}=0\,.
\ee
Since we are interested in supersymmetric solutions subject to the extremization equations \eqref{exme}, we will impose
the BPS equation \eqref{susye} in solving these equations. We emphasize that the extremization equations are derived before imposing the
supersymmetry constraint \eqref{susye}. We find that the first equation is solved by
\be
\label{extsol1}
\ell=\sqrt{P}\,,
\ee
which together with \eqref{susye} yields the solution to the second equation in \eqref{exme}
\be
\label{extsol2}
e_1=\frac14\sqrt{\frac{L_ H P+\l_1+\l_2}{G_3\ell q_1} }\,.
\ee
In turn, the solution for $e_1$ and the supersymmetry condition \eqref{susye} determines $e_2$. 
The solution to the third equation in \eqref{exme} yields
\be
\label{extsol3}
L_H=\frac{4G_3\ell^3q_2+\l_2 }{P}\,.
\ee
The last equation in \eqref{exme} is automatically satisfied once $\ell=P$ and the condition \eqref{susye}. Finally, substituting the above solutions for $\ell\,,e_1\,,e_2$ and $L_H$ into 
the entropy function \eqref{sfunction} 
we obtain 
\be
{\cal E}=\pi\sqrt{\frac{q_1\left(4G_3\ell^3q_2+\l_1+2\l_2\right)}{G_3\ell}}\,.
\ee
The expression above can be recast in a form similar to the Cardy formula 
\be
\label{entropy1}
{\cal E}=2\pi\sqrt{\frac{cq_1}{6}}\,,\quad c=\frac{3}{2G_3\ell}\left(4G_3\ell^3q_2+\l_1+2\l_2\right)\,.
\ee
From the dual 2D CFT point of view, the conserved charge $q_1$ is naturally interpreted as the effective energy level \cite{Peet1} inert under the higher derivative corrections \cite{deWit:2007dn}. Therefore it can be inferred from the known solution in the two-derivative theory using \eqref{extsol2}. The result is given by 
\be
\label{q1}
q_1=Q_2-\frac{\n^2}{PQ_1}\,,
\ee
in which the parameters $P$,$Q_1$, $Q_2$ and $\n$ are related to conserved charges carried by the rotating dyonic string and are integer valued in string units. The conserved charge $q_2$, however, does receive higher derivative corrections generically as it enters the central charge. Thus its value must be read off from the higher derivative corrected solutions. We will do so in the next section. The simplest way is via computing the horizon value of the dilaton and employing \eqref{extsol3}. 
 
From \eqref{3dansatz} and \eqref{extsol2}, we can
solve $r_+$ in terms of $q_1$ 
\be
r_+^2=\frac{4\ell^3G_3q_1}{L_H P +\l_1+\l_2}\,.
\ee
As shown in the next section, the value of $L_H$ increases when the higher derivative interactions are present. Therefore the size of the extremal BTZ horizon shrinks due to
higher derivative interactions, although the AdS$_3$ radius remains the same.

\section{The renormalized horizon value of dilaton}
This section is devoted to compute the value of dilaton curvature squared corrections, using which we can obtain the conserved charge $q_2$ from \eqref{extsol3}.
 For this purpose, the ansatz \eqref{genansatz} for a BPS string solution based on the U(2)$\times\mathbb{R}^2$ isometry is still applicable since 
the higher derivative extended field equations will not generate terms violating this symmetry. Supersymmetry requires the undetermined functions in \eqref{genansatz} to obey certain relations so that the corresponding Killing spinor equations 
\bea
\label{susyt}
0&=& (\partial_{\mu} +\frac14\omega_{\mu
	\a\b}\gamma^{\a\b})\epsilon^i+\frac18
H_{\mu\nu\rho}\gamma^{\nu\rho}\epsilon^i\, ,
\nn\\
0&=& \frac{1}{2\sqrt 2} \gamma^\mu
\delta^{ij}\partial_\mu L \epsilon_j  - \frac{1}{12\sqrt2}L\delta^{ij}\gamma_{\m\n\r} H^{\m\n\r} \epsilon_j \,, 
\eea
admit nontrivial solutions for the symplectic Majorana-Weyl spinor $\epsilon^i$. The indices $i$ and $j$ are raised and lowered by $\varepsilon^{ij}$ and $\varepsilon_{ij}$. We have also set auxiliary fields to 0 consistent with their field equations \cite{CP}. For convenience, we introduce the 
complex Weyl spinor 
\be
\epsilon=\epsilon_1+{\rm i}\epsilon_2\,, 
\ee
and assume the Killing spinor to have the form
\be
\label{ks1}
\epsilon=\Pi(r)\epsilon_0\,,
\ee 
where $\epsilon_0$ is the standard Killing spinor on a round 2-sphere embedded in the 6D spinor obeying the projection conditions
\be
\g^{012345}\epsilon_0=-\epsilon_0\,,\quad \g^{01}\epsilon_0=-\epsilon_0\,.
\ee
Plugging the ansatz for the bosonic fields and Killing spinor into \eqref{susyt},
we obtain the sufficient and necessary conditions for the existence of a Killing spinor:
\bea
\label{ks3}
0&=&A_0 \left(a_2^2 \left(c\varpi \dot{A}_0-c\dot{A}_3+2 A_3 b \right)-c\varpi \dot{d}\right)+c \left(a_1^2 \dot{\varpi}+A_3 \dot{d}-\dot{f}_1\right)+2 b\left(f_1-a_1^2 \varpi\right)\,,\\[0.1cm]
\label{ks4}
0&=&a_1 \left(a_2^2 \dot{A}_0+2 a_2 \left(A_0 \dot{a}_2+\dot{a}_1\right)+\dot{d}\right)+A_0 a_2 \left(\dot{d}-a_2^2 \dot{A}_0\right)\,,\\[0.1cm]
\label{ks5}
0&=&a_2^2 \left(c \left(\varpi \dot{A}_0-\dot{A}_3\right)+2 A_3 b\right)+2 b f_2-c \left(\dot{f}_2+\varpi \dot{d}\right)\,,\\[0.1cm]
\label{ks6}
0&=&-a_2^2 \dot{A}_0+2 a_1\dot{a}_2+\dot{d}\,,\\[0.1cm]
\label{ks7}
0&=&b \left(4 A_0 f_2 \varpi-4 A_3 f_2+c^2-4 f_1 \varpi-P \right)-c^2 \dot{c}\,,\\[0.1cm]
\label{ks8}
0&=&-a_2^3 A_3 \dot{A}_0+a_2\left(-A_0 \dot{f}_2+2 a_1 \left(\varpi \dot{a}_1+A_3 \dot{a}_2\right)+a_1^2 \dot{\varpi}+\dot{f}_1\right)+a_1 a_2^2 \dot{A}_3-a_1 \dot{f}_2\,,\\[0.1cm]
\label{ks9}
0&=&a_2 (f_1-A_0 f_2)+a_1 \left(a_2^2 A_3-f_2\right)+a_1^2 a_2 \varpi\,,\\[0.1cm]
\label{ks10}
0&=&\frac{8A_0 b f_2 \varpi}{c^3}-\frac{\dot{d}}{a_1 a_2}-\frac{8 A_3 b f_2}{c^3}-\frac{8 b f_1 \varpi}{c^3}-\frac{2 P  b}{c^3}+\frac{\dot{L}}{L}\,,\\[0.1cm]
0&=&\Pi(r)-\sqrt{a_1}\,,
\eea
where  dots denote derivatives with respect to $r$. The third last equation \eqref{ks9} is algebraic whose $r$ derivative is preserved on solutions to equations \eqref{ks4}, \eqref{ks6} and \eqref{ks8}. Equations. \eqref{ks4} and \eqref{ks6} together yield two first integrations 
\begin{equation}
\label{int1}
A_0+\frac{a_1}{a_2}=\k_1\,,\quad d+a_1a_2=\k_2\,,
\end{equation}
in which $\k_2$ can be set to 0 using the residual gauge symmetry of $B_{(2)}$. We can also choose $\k_1=0$ by shifting the $z$ coordinate and redefining
$f_2$ \footnote{For the rotating dyonic string solution without linear momentum \cite{CP}, $\k_1=1$ is the more common choice.}. Subsequently, 
applying the first equation in \eqref{int1} to  \eqref{ks8}, we find another first integration 
\be
\label{int2}
a_1a_2A_3+a_1^2\varpi+f_1=\k_3\,.
\ee
The equation above together with \eqref{ks9} forces $\k_3=0$.
From \eqref{ks7} and \eqref{ks10}, we also find
\be
\label{solb}
|d|Lc^2=r^2\,,
\ee
after choosing the $r$ coordinate so that $c=br$.
Using \eqref{int1} and \eqref{int2}, we obtain two more first integrations from Eqs. \eqref{ks3} and \eqref{ks5}
\be
\label{int3}
f_2+d\varpi=r^2\k_4\,,\quad A_3=A_0\varpi+r^2\k_5\,,
\ee
where $\k_4$ and $\k_5$ will be set to 0 in order for the solution to obey the asymptotically flat boundary condition. Substituting \eqref{int3} to Eqs.\eqref{int2} and \eqref{ks7}, we obtain
\be
\label{solc}
f_1=0\,,\quad c^2=r^2+P\,.
\ee
In summary, the off-shell Killing spinor equations for the asymptotically flat rotating dyonic string ansatz 
are fully solved provided the following relations are satisfied
\bea
\label{kstot}
A_0+\frac{a_1}{a_2}&=&0\,,\quad d+a_1a_2=0\,,\quad c^2=r^2+P\,,\quad c=br\,,\nn\\
|d|Lc^2&=&r^2\,,\quad f_1=0\,,\quad f_2=-d\varpi\,,\quad A_3=A_0\varpi\,,
\eea
which indicates that the horizon value of $L$ is determined by the near-horizon 
behavior of $d$. This is a significant simplification as the $d$ equation is associated with a conservation law
and already admits a first integration. 

In terms of the new radial coordinate 
$\r=r^2$ and applying the set of relations \eqref{kstot}, we read off the $d$ equation from the $(t,z)$ component of the $B_{(2)}$ field equation. It turns out to be independent of other undetermined variables in the ansatz and can be integrated twice so that the resulting equation is only second order in $\r$ derivative. Imposing the asymptotically flat boundary condition, 
one can set one of the integration constants to 0 and the $d$ equation reads 
\be
\left(\r-(\r+Q_1)d\right)d+\frac{(\l_2-\l_1)P^2}{2(P+\r)^2}d^2
-\frac{2\r\l_1+(\l_1+\l_2)P}{P+\r}\r dd'+\l_1\r^2d'^2-\l_1\r^2dd''=0\,,
\ee
where the prime denotes the $\r$ derivative. The simple form of the $d$ equation has benefited from
the precise structure of the off-shell supersymmetric actions. Had we performed a field redefinition, 
this simplification would be immediately lost . In the context of string compactification, $\l_1$ and $\l_2$
are of the same order much smaller than $P$ or $Q_1$. The equation above can thus be solved order by order algebraically in the small parameters $\l$.
Here, only the corrections at the first order in $\l$ are meaningful, because higher-order terms will be modified by possible $R^n\,(n>2)$ interactions which are not included in our discussion. To the first order in $\l$, the solution of the $d$ equation is given by
\be
d(\r)=\frac{\r}{\r+Q_1}\left[1-\frac{P^2Q_1^2(\l_1+\l_2)+2PQ_1^2(\l_1+\l_2)\r+(Q_1^2(\l_1+\l_2)+(P-Q_1)^2(\l_1-\l_2))\r^2}{2(P+\r)^2(Q_1+\r)^3}\right]\,,
\ee
from which one can see that corrections to the leading-order solution stay small within the entire range of $\r$ as long as 
$\l$s are much less than $P$ or $Q_1$. 
In the special case $\l_1=\l_2$, the combination of the curvature squared superinvariants is consistent with 16 supercharges, and the solution is independent of $P$. Meanwhile the conserved charge $q_2$ \eqref{extsol3} is not modified by the higher derivative interactions, which is physically more reasonable as the curvature squared terms do not modify the conserved charges carried by the asymptotically flat dyonic string solution. The equal charge case $P=Q_1$ corresponds to a BMPV black hole \cite{BMPV}, and the solution also looks quite simple depending only on the sum of $\l$'s. Near the string horizon at $\r=0$, 
\be
d(\r)=\frac{\r}{Q_1}\left(1-\frac{\l_1+\l_2}{2Q_1}\right)+{\cal O}(\r^2)\,.
\ee
Combining this result with the BPS equation relating $d$ and $L$, we obtain the horizon value of the dilaton with curvature squared corrections 
\be
L_H=\frac{Q_1+\ft12(\l_1+\l_2)}P\Rightarrow q_2=\frac{1}{4G_3\ell^3}\left(Q_1+\ft12(\l_1-\l_2)\right)\,.
\ee
Substituting the result above to \eqref{entropy1}, we complete the calculation of the macroscopic entropy for the higher derivative corrected rotating BPS dyonic string solution. The final entropy formula takes the form
\be
\label{entropy2}
{\cal E}=2\pi\sqrt{\frac{cq_1}{6}}\,,\quad c=\frac{3}{2G_3\ell}\left(Q_1 +\ft32(\l_1+\l_2)\right)\,,\quad q_1=Q_2-\frac{\n^2}{PQ_1}\,.
\ee
The sum of $\l$s is related to the total coefficient in front of the $B_{(2)}\wedge {\rm Tr}(R\wedge R)$ term or the Riemann tensor squared. It also means each independent supersymmetric completion of the Riemann tensor squared contributes equally to the entropy. 
In the IIA string embedding, $\l_1+\l_2$ is given by $2\a'$ \cite{LM}. Since entropy is a dimensionless quantity, it does not depend on the choice of units. In particular,
we can choose the string units in which the string length $\ell_s=1$. Also, we recall from \cite{CP} that in the string units $1/{G_3}=4P^{\frac{3}2}$. Therefore the central charge is equal to $c=6P(Q_1+3)$ indicating that the macroscopic entropy with leading-order higher derivative correction is of the form
\be
{\cal E}=2\pi\sqrt{PQ_1Q_2-\n^2}\left(1+\frac3{2Q_1}\right)\,,
\ee
which is consistent with the result obtained in \cite{dK} specialized to the K3$\times T^2$ compactification. 
One should also be aware that in the string units, the parameters $P,\,Q_1$, $Q_2$ and $\n$ are all integer valued \cite{CY}. The results obtained here and \cite{dK} agree with the one reported in \cite{CDKL} only in the static case $\n=0$. 
It should be interesting to connect the macroscopic result to the microscopic one in certain asymptotic limits \cite{CM, Banerjee:2008ag}.
\section{Conclusion and discussions}

In this work, we computed the macroscopic entropy of the supersymmetric rotating dyonic strings in 6D (1,0) supergravity with curvature squared corrections, adopting Sen's entropy function formalism applied to the near-horizon geometry of the string. Upon solving the attractor equations whose solution extremizes  
the entropy function, we observed that inclusion of the higher derivative interactions does not modify the $AdS_3$ radius of the BTZ black hole but does shrink the size of the extremal horizon. Expressing the extremized entropy function in terms of the conserved charges carried by the string requires the knowledge of the horizon value of the dilaton  which can be extracted only from the complete asymptotically flat string solution.  We also showed that although there are two independent supersymmetric completions of the Riemann tensor squared, the entropy is insensitive to  their detailed structures but depends only on the coefficient of the Riemann squared term fixed to be $\a'/8$ upon embedding the 6D supergravity model in the K3 compactification of IIA string.

Via a circle reduction, the rotating dyonic string becomes the three-charge spinning black hole in the 5D ungauged STU model. Thus our result can be compared to previous work \cite{CDKL,dK} on the macroscopic entropy of 
supersymmetric black holes in 5D ${\cal N}=2$ supergravity with curvature squared corrections descending from M theory compactified on K3$\times T^2$. Our entropy formula is in agreement with the result obtained in \cite{dK} but exhibits a different angular momentum dependence from the result reported in \cite{CDKL}. In the future, we hope to compute the microscopic entropy of the dual CFT$_2$ and compare it to the macroscopic result. 

String dualities relate IIA string on K3 to heterotic string on $T^4$ in which the black hole entropy with $\a'$ corrections was previous obtained in \cite{Sahoo:2006pm,CMOR}\footnote{In the recent work \cite{CMOR} revisiting the $\a'$-corrected black hole in heterotic string, the coefficient in front of the supersymmetric black hole should be $\a'/8$ instead of $\a'$.}. A comparison between the entropy of black holes in the heterotic and IIA strings shows that once $\a'$ corrections are taken into account, the black hole entropy is not invariant under the duality transformation.  

Locally the extremal BTZ$\times S^3$ is equivalent to $AdS_3\times S^3$ around which the spectrum of fluctuations has been analyzed in \cite{NOPT,BNOPT}. The fact that the spectrum can be organized into representations of SU(1,1$| 2)\times$SU$(1,1)\times $SU(2) \cite{NOPT,BNOPT} suggests that the $S^3$ compactification of the 6D model can be described by the matter coupled 3D supergravity model based on the SU$(1,1 |2)\times$SU(1,1) algebra. Omitting the curvature squared terms, the action for the massless fields in the supergravity multiplet can be formulated as a Chern-Simons gauge theory. After integrating out auxiliary fields, it takes the form \cite{AT,sen1,GTW},
\bea
S_{3d}&=&\int d^3x\left[k_G\sqrt{-g}\left(R+\frac2{\ell^2}\right)+k_{\rm CS}\Omega(\G)-(\ell k_G-k_{\rm CS})\Omega(A_R)\right]\,,\nn\\
\Omega(\G)&=&\ft12{\rm Tr}(\G d\G+\ft23\G^3)\,,\quad \Omega(A_R)={\rm Tr}(A_R dA_R+\ft23A_R^3)\,,
\eea
where $A_R$ is the gauge field associated with the SU(2) inside SU(1,1$| 2)$ and various constants are given by
\be
k_G=\frac{PL_{H}+\a'}{16\pi G_3P}=\frac{\sqrt{P}(Q_1+2)}{4\pi}\,,\quad k_{\rm CS}=-\frac{\a'}{16\pi G_3\sqrt{P}}=-\frac{P}{4\pi}\,.
\ee
One feature is that the effective Newton's constant depends on the value of dilaton at the string horizon that is affected by the higher derivative interaction via equations of motion. The level of the SU(2)$_R$ Chern-Simons gauge field is proportional to the central charge appearing in the entropy formula
\be
k_R=-\frac{P(Q_1+3)}{4\pi}\,.
\ee
The fact that the quantity above obeys the  quantization condition $4\pi k_R=n\,,n=0,\pm 1\cdots$ \cite{Deser:1982vy} implies that the higher-derivatice corrections to the horizon value of $L$ can be at most quadratic in $\a'$. For instance, $\a'^2/P^2$ may contribute a $P$-independent integer to $4\pi k_R$. Based on dimension analysis, such a term would come from a supersymmetric $R^3$ action which is forbidden in both heterotic string and type II string \cite{Bd}. Therefore the value of dilation on the string horizon does not receive corrections from higher derivative interactions beyond order $\a'$.

\section*{Acknowledgements}  
I am grateful to David Chow for joining the early stage of this work. I also thank Istanbul Technical University and Tianjin University for hospitality during my visit. The work of Y.P. is supported by a Newton International Fellowship No. NF170385 of Royal Society.


\begin{thebibliography}{99}

\bibitem{SV1}
  A.~Strominger and C.~Vafa,
  ``Microscopic origin of the Bekenstein-Hawking entropy,''
  Phys.\ Lett.\ B {\bf 379} (1996) 99, hep-th/9601029.

\bibitem{BMPV}
  J.~C.~Breckenridge, R.~C.~Myers, A.~W.~Peet and C.~Vafa,
  ``D-branes and spinning black holes,''
  Phys.\ Lett.\ B {\bf 391} (1997) 93, hep-th/9602065.
  
\bibitem{MSW}
  J.~M.~Maldacena, A.~Strominger and E.~Witten,
  ``Black hole entropy in M theory,''
  JHEP {\bf 9712} (1997) 002, hep-th/9711053.
  

\bibitem{Peet1}
  A.~W.~Peet,
  ``TASI lectures on black holes in string theory,''
   hep-th/0008241.  

\bibitem{David:2002wn}
  J.~R.~David, G.~Mandal and S.~R.~Wadia,
  ``Microscopic formulation of black holes in string theory,''
  Phys.\ Rept.\  {\bf 369} (2002) 549, hep-th/0203048.
  
\bibitem{sen1}
  A.~Sen,
  ``Black Hole Entropy Function, Attractors and Precision Counting of Microstates,''
  Gen.\ Rel.\ Grav.\  {\bf 40} (2008) 2249, arXiv:0708.1270 [hep-th].
  
\bibitem{Mandalsen}
  I.~Mandal and A.~Sen,
  ``Black Hole Microstate Counting and its Macroscopic Counterpart,''
  Nucl.\ Phys.\ Proc.\ Suppl.\  {\bf 216} (2011) 147
   [Class.\ Quant.\ Grav.\  {\bf 27} (2010) 214003], arXiv:1008.3801 [hep-th].

\bibitem{Sen3}
  A.~Sen,
  ``Microscopic and Macroscopic Entropy of Extremal Black Holes in String Theory,''
  Gen.\ Rel.\ Grav.\  {\bf 46} (2014) 1711, arXiv:1402.0109 [hep-th]. 

\bibitem{CdM1}
  G.~Lopes Cardoso, B.~de Wit and T.~Mohaupt,
  ``Corrections to macroscopic supersymmetric black hole entropy,''
  Phys.\ Lett.\ B {\bf 451} (1999) 309, hep-th/9812082.
  
\bibitem{CdM2}
  G.~Lopes Cardoso, B.~de Wit and T.~Mohaupt,
  ``Deviations from the area law for supersymmetric black holes,''
  Fortsch.\ Phys.\  {\bf 48} (2000) 49, hep-th/9904005.
 
\bibitem{Vafa:1997gr}
  C.~Vafa,
  ``Black holes and Calabi-Yau threefolds,''
  Adv.\ Theor.\ Math.\ Phys.\  {\bf 2} (1998) 207,
  hep-th/9711067.
   

\bibitem{CDKL}
  A.~Castro, J.~L.~Davis, P.~Kraus and F.~Larsen,
  ``Precision Entropy of Spinning Black Holes,''
  JHEP {\bf 0709} (2007) 003, arXiv:0705.1847 [hep-th].
  
\bibitem{dK}
  B.~de Wit and S.~Katmadas,
  ``Near-Horizon Analysis of D=5 BPS Black Holes and Rings,''
  JHEP {\bf 1002} (2010) 056, arXiv:0910.4907 [hep-th]. 
  
  
\bibitem{LM}
  J.~T.~Liu and R.~Minasian,
  ``Higher-derivative couplings in string theory: dualities and the B-field,''
  Nucl.\ Phys.\ B {\bf 874} (2013) 413.

 \bibitem{Bergshoeff:1985mz}
  E.~Bergshoeff, E.~Sezgin and A.~Van Proeyen,
  ``Superconformal Tensor Calculus and Matter Couplings in Six-dimensions,''
  Nucl.\ Phys.\ B {\bf 264} (1986) 653, Erratum: [Nucl.\ Phys.\ B {\bf 598} (2001) 667]. 


 \bibitem{NOPT}
  J.~Novak, M.~Ozkan, Y.~Pang and G.~Tartaglino-Mazzucchelli,
  ``Gauss-Bonnet supergravity in six dimensions,''
  Phys.\ Rev.\ Lett.\  {\bf 119} (2017) no.11,  111602.
 
\bibitem{BNOPT}
  D.~Butter, J.~Novak, M.~Ozkan, Y.~Pang and G.~Tartaglino-Mazzucchelli,
  ``Curvature squared invariants in six-dimensional ${\cal N} = (1,0)$ supergravity,''
  JHEP {\bf 1904} (2019) 013, 
  arXiv:1808.00459 [hep-th].
  
\bibitem{BRS}
  E.~A.~Bergshoeff, J.~Rosseel and E.~Sezgin,
  ``Off-shell D=5, N=2 Riemann Squared Supergravity,''
  Class.\ Quant.\ Grav.\  {\bf 28} (2011) 225016, arXiv:1107.2825 [hep-th].
  

 
  
\bibitem{OP1}
  M.~Ozkan and Y.~Pang,
  ``Supersymmetric Completion of Gauss-Bonnet Combination in Five Dimensions,''
  JHEP {\bf 1303} (2013) 158
   Erratum: [JHEP {\bf 1307} (2013) 152],
  arXiv:1301.6622 [hep-th].
  
\bibitem{OP2}
  M.~Ozkan and Y.~Pang,
  ``All off-shell $R^{2}$ invariants in five dimensional $\mathcal{N} =$ 2 supergravity,''
  JHEP {\bf 1308} (2013) 042, arXiv:1306.1540 [hep-th].
 


\bibitem{HOT}
  K.~Hanaki, K.~Ohashi and Y.~Tachikawa,
  ``Supersymmetric Completion of an R**2 term in Five-dimensional Supergravity,''
  Prog.\ Theor.\ Phys.\  {\bf 117} (2007) 533, hep-th/0611329.

\bibitem{CL}
  M.~Cveti\v{c} and F.~Larsen,
  ``Near horizon geometry of rotating black holes in five dimensions,''
  Nucl.\ Phys.\ B {\bf 531}, 239  (1998), hep-th/9805097.

 
 \bibitem{CY}
  M.~Cvetic and D.~Youm,
  ``General rotating five-dimensional black holes of toroidally compactified heterotic string,''
  Nucl.\ Phys.\ B {\bf 476} (1996) 118.
  
\bibitem{CC}
  D.~D.~K.~Chow and G.~Compère,
  ``Black holes in N=8 supergravity from SO(4,4) hidden symmetries,''
  Phys.\ Rev.\ D {\bf 90} (2014) no.2,  025029, arXiv:1404.2602 [hep-th].
  
  
\bibitem{CP}
  D.~D.~K.~Chow and Y.~Pang,
  ``Rotating Strings in Six-Dimensional Higher-Derivative Supergravity,''
  arXiv:1906.07426 [hep-th].



  
\bibitem{BSS86} 
 E.~Bergshoeff, A.~Salam and E.~Sezgin,
  ``A Supersymmetric R**2 Action in Six-dimensions and Torsion,''
 Phys.\ Lett.\ B {\bf 173}, 73 (1986).
 
 
 \bibitem{O} 
M.~Ozkan, “Supersymmetric curvature squared invariants in five and six dimensions,” PhD Thesis, Texas A\& M University, 2013.

 
  
\bibitem{GMR}
  J.~B.~Gutowski, D.~Martelli and H.~S.~Reall,
  ``All Supersymmetric solutions of minimal supergravity in six- dimensions,''
  Class.\ Quant.\ Grav.\  {\bf 20} (2003) 5049.
  
\bibitem{AP}
  M.~Akyol and G.~Papadopoulos,
  ``Topology and geometry of 6-dimensional (1,0) supergravity black hole horizons,''
  Class.\ Quant.\ Grav.\  {\bf 29} (2012) 055002, arXiv:1109.4254 [hep-th].

\bibitem{KuLu}
  H.~K.~Kunduri and J.~Lucietti,
  ``Classification of near-horizon geometries of extremal black holes,''
  Living Rev.\ Rel.\  {\bf 16} (2013) 8, arXiv:1306.2517 [hep-th].
  
\bibitem{Deger:2014ofa}
  N.~S.~Deger, H.~Samtleben, O.~Sarioglu and D.~Van den Bleeken,
  ``A supersymmetric reduction on the three-sphere,''
  Nucl.\ Phys.\ B {\bf 890} (2014) 350,
  arXiv:1410.7168 [hep-th].
  
\bibitem{Deger:2019jtl}
  N.~S.~Deger, N.~Petri and D.~Van den Bleeken,
  ``Supersymmetric Dyonic Strings in 6-Dimensions from 3-Dimensions,''
  JHEP {\bf 1904} (2019) 168,
 arXiv:1902.05325 [hep-th].

  
\bibitem{SS}
  B.~Sahoo and A.~Sen,
  ``BTZ black hole with Chern-Simons and higher derivative terms,''
  JHEP {\bf 0607} (2006) 008, hep-th/0601228.

 \bibitem{deWit:2007dn}
  B.~de Wit,
  ``BPS black holes,''
  Nucl.\ Phys.\ Proc.\ Suppl.\  {\bf 171} (2007) 16,arXiv:0704.1452 [hep-th]. 
 

\bibitem{AT} 
A.~Achucarro and P.~K.~Townsend, 
“A Chern-Simons Action For Three- Dimensional Anti-De Sitter Supergravity Theories,” 
Phys.\ Lett.\ B\ {\bf 180}, 89 (1986).
  
\bibitem{GTW}
  A.~Giacomini, R.~Troncoso and S.~Willison,
  ``Three-dimensional supergravity reloaded,''
  Class.\ Quant.\ Grav.\  {\bf 24} (2007) 2845,
  hep-th/0610077.
 
 


\bibitem{CMOR}
  P.~A.~Cano, P.~Meessen, T.~Ortin and P.~F.~Ramírez,
  ``$\alpha'$-corrected black holes in String Theory,''
  JHEP {\bf 1805} (2018) 110, arXiv:1803.01919 [hep-th].
  
  
\bibitem{Sahoo:2006pm}
  B.~Sahoo and A.~Sen,
  ``alpha-prime - corrections to extremal dyonic black holes in heterotic string theory,''
  JHEP {\bf 0701} (2007) 010, hep-th/0608182.


\bibitem{CM}
  A.~Castro and S.~Murthy,
  ``Corrections to the statistical entropy of five dimensional black holes,''
  JHEP {\bf 0906} (2009) 024, arXiv:0807.0237 [hep-th].
  
\bibitem{Banerjee:2008ag}
  N.~Banerjee,
  ``Subleading Correction to Statistical Entropy for BMPV Black Hole,'' Phys.\ Rev.\ D {\bf 79} (2009) 081501, arXiv:0807.1314 [hep-th]. 
  
\bibitem{Deser:1982vy}
  S.~Deser, R.~Jackiw and S.~Templeton,
  ``Three-Dimensional Massive Gauge Theories,''
  Phys.\ Rev.\ Lett.\  {\bf 48} (1982) 975.
  
\bibitem{Bd}
  E.~A.~Bergshoeff and M.~de Roo,
  ``The Quartic Effective Action of the Heterotic String and Supersymmetry,''
  Nucl.\ Phys.\ B {\bf 328} (1989) 439.
  


 
  
\end{thebibliography}

\end{document}